\documentclass[preprint,superscriptaddress,eqsecnum,nofootinbib]{revtex4}
\usepackage{graphicx,amssymb,amsbsy,bm,amsmath,psfrag,dsfont}
\usepackage{slashed}
\usepackage{hyperref}
\usepackage[dvips]{color}

\begin{document}

\newcommand{\be}{\begin{equation}}
\newcommand{\ee}{\end{equation}}
\newcommand{\bea}{\begin{eqnarray}}
\newcommand{\eea}{\end{eqnarray}}
\newcommand{\no}{\noindent}

\newcommand{\la}{\lambda}
\newcommand{\si}{\sigma}
\newcommand{\vk}{\vec{k}}
\newcommand{\vx}{\vec{x}}
\newcommand{\om}{\omega}
\newcommand{\Om}{\Omega}
\newcommand{\ga}{\gamma}
\newcommand{\Ga}{\Gamma}
\newcommand{\gaa}{\Gamma_a}
\newcommand{\al}{\alpha}
\newcommand{\ep}{\epsilon}
\newcommand{\app}{\approx}
\newcommand{\uvk}{\widehat{\bf{k}}}
\newcommand{\OM}{\overline{M}}

\title{Inflatonless Inflation}
\author{Chiu Man Ho} \email{chiuman.ho@vanderbilt.edu}
  \affiliation{Department of
  Physics and Astronomy, Vanderbilt University, Nashville, TN 37235, USA}
\author{Thomas W. Kephart}\email{tom.kephart@gmail.com}
\affiliation{Department of
  Physics and Astronomy, Vanderbilt University, Nashville, TN 37235, USA}

\date{\today}

\begin{abstract}
We consider a $4+N$ dimensional Einstein gravity coupled to a non-linear sigma model.
This theory admits a solution in which the $N$ extra dimensions contract exponentially while the ordinary space
expand exponentially. Physically, the non-linear sigma fields induce the
dynamical compactification of the extra dimensions, which in turn drives inflation. No inflatons are required.
\end{abstract}

\maketitle

\section{Introduction}


Inflation has been the most convincing scheme to solve the horizon, flatness and monopole problems \cite{Turner,Dodelson}.
Usually, theoretical models of inflation involve some kind of scalar fields called inflatons, whose potential energy dominates
over the kinetic energy during the phase of accelerated expansion \cite{Mazumdar}. The stage right after inflation is
called reheating \cite{Linde,Reheat} at which these inflatons decay into all other observed particles. As pointed out in
\cite{Brandenberger}, inflationary models involving inflatons generally suffer from issues of fine-tuning and initial
singularity. This appears to call for the exploration of possible alternatives to the conventional inflation schemes
\cite{Brandenberger,Gratton,TachyonAlternatives,Dutta}. However, it is fair enough to say that the standard scalar
field inflation is the best scenario proposed so far. Despite this fact, we think that it is still important to look for
other alternatives. For instance, a direct question could be: Can inflation happen without inflatons?  This question was also addressed in \cite{Watson} by assuming that the universe has undergone cascading energy transitions, but these authors did not provided a specific model
that realized the assumption.

On the other hand, superstring theory \cite{PolchinskiBook} proposes that we are actually living in a ten-dimensional spacetime,
six of which are compactified. The natural size of these compactified extra dimensions is expected to be of the same order as the
Planck length. This explains why we have never observed them at the energy scale that we have been probing. But at the same time, this
immediately poses a fundamental question: Does the compactification of these extra dimensions have a dynamical origin? Some interesting
works concerning this issue include \cite{Goldberger,Zhuk,Chethan}.

The purpose of this article is to search for a \emph{simultaneous} solution to both of the above questions. Similar attempts by invoking a
$4+N$ dimensional Einstein gravity with or without matter sources, where $N$ is the total
number of extra spatial dimensions, have been conducted in \cite{Alvarez:1983kt,Sahdev:1988fp,Kolb:1984ga,Levin:1994yw}.
However, a common problem associated with these schemes is that they cannot lead to an exponential expansion of the
ordinary space.

In this article, we will consider a $4+N$ dimensional Einstein gravity coupled to
a non-linear sigma model. A similar model was first proposed in \cite{Omero:1980vx,GellMann:1984sj,GellMann:1985if} to induce
a spontaneous compactification of the extra dimensions. It was then utilized to induce both compactification and ``inflation''
in \cite{Chakrabarty:1989uf}. But again the expansion of the ordinary space is not exponential. In order to resolve this problem,
we will perform a simple extension to the original model. As a consequence, we are able to find a cosmological solution where
exponential inflation and dynamical compactification of the extra dimensions occur simultaneously.

In fact, since the work of \cite{Levin:1994yw,Omero:1980vx,GellMann:1984sj,GellMann:1985if}, the idea of the connection between inflation and dynamical compactification has been continually explored, particularly in string inflation models as reviewed in \cite{Silverstein}. However, what we have set out to achieve is \emph{not} to improve any of these string inflation models which contain inflatons in our 4D spacetime in their low-energy effective Lagrangian. The main achievement of our article is to provide a model in which inflation happens \emph{without inflatons}, but instead is solely driven by the dynamical compactification of extra dimensions. This was precisely what \cite{Levin:1994yw,Omero:1980vx,GellMann:1984sj,GellMann:1985if} had hoped to achieve.

The article is organized as follows. In Section II, we will construct our model and obtain the equations of motion.
In Section III, we  obtain a simple solution in which the extra dimensions
contract exponentially while the ordinary space expand exponentially. Finally, we will draw our conclusions and include some
discussions in Section IV.

\section{The Model}

Before getting into the dynamics by specifying an action for our model, we will first lay out the geometric part of the Einstein
field equations. We start with a generalized Friedmann-Robertson-Walker (FRW) diagonal metric in a $4+N$ spacetime:
\bea
\label{metric}
ds^2 = dt^2- a^2(t) \,g_{i j} \,dx^{i} \,dx^{j}-b^2(t)\, g_{mn}\,dy^{m}\,dy^n \,,
\eea
where $i,j=1,2,3$ and $m,n=5,6, ..., \,4+N$, \,$a(t)$ and $b(t)$ are the scale factors associated with the ordinary and extra (spatial)
dimensions respectively. For instance, we have $N=6$ in superstring theory. However, to retain the generality of our study, we will keep $N$ unspecified. The various components of the Ricci tensor $R_{AB}$
resulting from the above metric (\ref{metric}) are
\bea
\label{R0}
R_{00} &=& -3\,\frac{\ddot{a}(t)}{a(t)}-N\,\frac{\ddot{b}(t)}{b(t)}\,, \\
\label{Ri}
R_{ij} &=& a^2(t)\, g_{ij}\,\Bigg\{\, \dot{H}_a(t)+\left[\,3 \, H_a(t)+ N \,H_b(t)\,\right] \, H_a(t)\,\Bigg\}\,, \\
\label{Rm}
R_{mn} &=& b^2(t)\, g_{mn}\,\Bigg\{\, \dot{H}_b(t)+ \left[\,3 \, H_a(t)+ N \,H_b(t)\,\right] \, H_b(t)+\frac{2\, K}{b^2(t)} \,\Bigg\}\,,
\eea
where $K$ is the possible constant curvature of the compact extra dimensions and
\bea
H_a(t)=\frac{\dot{a}(t)}{a(t)} ~~;~~ H_b(t)=\frac{\dot{b}(t)}{b(t)}\,.
\eea

As anticipated, our theory is an extension of the model proposed by \cite{Omero:1980vx,GellMann:1984sj,GellMann:1985if,Chakrabarty:1989uf}
in which the $4+N$ dimensional Einstein gravity is coupled to a non-linear sigma model. Our $4+N$ dimensional action is given by
\bea
\label{action}
S=\int\,d^{4+N}Z\, \sqrt{G} \left[\,-\frac12\, M^{2+N} \,R +\frac{1}{\la^2}\,
G^{AB}\,h_{CD}(\phi)\,\partial_{A}\,\phi^C\,\partial_B\,\phi^D+ g \,G^{AB}\,h_{AB}(\phi)\,(\partial_{C}\,\phi^C)^2\,\right]\,,
\nonumber \\
\eea
where $Z=\{t,\, x^i,\,y^m\}$,\, $G_{AB}$ and $R$ are the metric and the Ricci scalar of the $4+N$ dimensional spacetime respectively,
$1/\lambda^2$ and $g$ are coupling constants, $M$ is the $(4+N)$ dimensional Planck mass, $\phi^C$ are the non-linear sigma fields, and the
target space manifold metric $h_{CD}$ determines
the dynamics of the non-linear sigma fields. We choose the number of non-linear sigma fields to be exactly the same
as the dimension of the spacetime. Thus, throughout the entire article, both of the spacetime and target-space indices
run from $1$ to $4+N$. We note in passing that this choice for the number of non-linear sigma fields is crucial for the
feasibility of our model.

The last term in the action, \, $g \,G^{AB}\,h_{AB}(\phi)\,(\partial_{C}\,\phi^C)^2$,\, is a new term
that we have added to the original model in \cite{Omero:1980vx,GellMann:1984sj,GellMann:1985if,Chakrabarty:1989uf}. As we will
see, it plays a crucial role in giving the simultaneous solution to inflation and dynamical compactification.
However, this new term in the action is not generally covariant. It is because the spacetime and target-space indices have been
contracted in a mixed way.  In order to make the action covariant, we begin by imposing the following ansatz:
\bea
\label{ansatz1}
\phi^{0} &=& \;\beta\;\; t ~~~~\textrm{with}~~~~ \beta =0 \,,\\
\label{ansatz2}
\phi^{i} &=& \beta'\, x^{i} ~~~~\textrm{with}~~~~ \beta'=0\,,\\
\label{ansatz3}
\phi^{m} &=& \alpha \, y^{m} + \; \textrm{constant}\,,
\eea
where $\al$ is a characteristic constant carrying the mass dimension of $\frac{4+N}{2}$. This is precisely the same ansatz that
was imposed by \cite{Omero:1980vx,GellMann:1984sj,GellMann:1985if,Chakrabarty:1989uf} to induce dynamical compactification
and hence inflation, although they failed to generate an exponential inflation. Interestingly, 't Hooft made a similar ansatz
in \cite{Hooft} for other reasons. Physically, this ansatz can be understood as a particular set of vacuum expectation values
acquired by the non-linear sigma fields. Another way to understand this is that we have identified
some components of the non-linear sigma fields with the coordinates of the extra dimensions. This is intuitively conceivable
because the non-linear sigma fields themselves are ``coordinates" of the target space manifold. Mathematically, this ansatz is also justified as
the non-linear sigma fields are functions of the spacetime coordinates, and it simply specifies an explicit dependence of the non-linear
sigma fields on the spacetime coordinates.

The above ansatz also implies that the non-linear sigma fields are totally independent of the ordinary 4D spacetime coordinates.
Thus, they only live in the extra dimensions. Except for the metric
associated the ordinary space, everything in the theory is independent of the ordinary 4D spacetime coordinates.
Note that the ``constant" appearing
in \eqref{ansatz3} has been set to zero in \cite{Omero:1980vx,GellMann:1984sj,GellMann:1985if,Chakrabarty:1989uf}.
While we keep it for generality, none of the results in
our work or in \cite{Omero:1980vx,GellMann:1984sj,GellMann:1985if,Chakrabarty:1989uf} depend on it.

In light of the above ansatz, the spacetime indices coincide
exactly with the target-space indices. Now, the action becomes
\bea
\label{actionCovariant}
S'=\int\,d^{4+N}Z\, \sqrt{G} \left[\,-\frac12\, M^{2+N} \,R +\frac{\alpha^2}{\la^2}\,
G^{AB}\,h_{mn}\,\Delta_{Am}\,\Delta_{Bn}+ g\,N^2 \,\alpha^2\,G^{AB}\,h_{AB}\,\right]\,,
\eea
where we have defined
\bea
\Delta_{Am}=\left\{
              \begin{array}{ll}
                \delta_{Am}, & \hbox{\textrm{~~for}~ $A \,=\, 5,\,6,\, ..., \,4+N$\,;} \\
                0, & \hbox{\textrm{~~for}~ $A\; = \; 0,\,1,\,2,\,3$\,.}
              \end{array}
            \right.
\eea
Originally, $h_{AB}(\phi)$ was a tensor in
the target space and depended on the non-linear sigma fields.
Once the ansatz is imposed, $h_{AB} $ is solely dependent on the spacetime coordinates. However, the
operation of the ansatz, which is mathematically consistent, should not affect the tensor identity of
$h_{AB}$. Thus, we expect the components of $h_{AB}$ to be proportional to components of $G_{AB}$. This will be
confirmed later on when we solve for the actual expressions for $h_{AB}$.  Also, there is now no more
uncontracted spacetime indices or target-space indices in $S'$. We thus conclude that general covariance has
emerged when the non-linear sigma fields obey the identifications \eqref{ansatz1}, \eqref{ansatz2} and \eqref{ansatz3}.
As a result, the action $S'$ is generally covariant.

One may wonder about the origin of the general covariance emerged in \eqref{actionCovariant}.
We can understand this as follows.
Originally, we have two different sets of diffeomorphisms. One
is in the internal space and the other is the non-compact spacetime.
Let's call their diffeomorphism algebras $D_1$ and $D_2$
respectively. Before the two are coupled  we have $D_1 \otimes D_2$
as the symmetry algebra of the theory. Once they are coupled,
only the diagonal subalgebra
$D \subset (\,D_1 \otimes D_2\,)$ is preserved. The purpose of the ansatz in \eqref{ansatz1}, \eqref{ansatz2} and \eqref{ansatz3} is precisely
to extract the diagonal subalgebra, which leads to the general covariance in \eqref{actionCovariant}. In this respect,
it appears that the situation is analogous to, for example,
chiral symmetry breaking in QCD where the groups $SU(N_f)_L$ and $SU(N_f)_R$ of the
underlying QCD are spontaneously broken down to diagonal subgroup $SU(N_f)_V$. The resulting mass
term of the fermions has precisely the same type of interactions as above if we associate left-handed fermions with
scalar fields and right-handed fermions with spacetime metric. Finally, we note that the idea of ``soldering" internal and external indices has been
used by Polyakov in the context of non-critical string theory \cite{polyakov}, where it is argued
that general covariance emerges from the underlying (gauge) current algebra structure.

\section{Inflation from Dynamical Compactification}

To iterate, our study is based on the action $S'$ which is generally covariant. The equations of motion following from the
variation of the action $S'$ with respect to the metric $G^{AB}$ is given by
\bea
\label{RAB}
R_{AB}=\frac{2\,\alpha^2}{M^{2+N}}\left(\, \frac{1}{\la^2}\,h_{mn}\,\Delta_{Am}\,\Delta_{Bn} +
g\,N^2\, h_{AB}  \,\right)\,,
\eea
which leads to following components
\bea
\label{R00}
R_{00}&=&\frac{2\,\alpha^2}{M^{2+N}}\,N^2\, g \, h_{00} \,,\\
\label{Rij}
R_{ij}&=&\frac{2\,\alpha^2}{M^{2+N}}\,N^2\, g \, h_{ij} \,,\\
\label{Rmn}
R_{mn}&=&\frac{2\,\alpha^2}{M^{2+N}} \left(\,\frac{1}{\la^2}+N^2\, g \,\right) h_{mn}\,.
\eea
It is clear that if $g=0$, then $R_{00}=R_{ij}=0$, which is in agreement with \cite{Omero:1980vx,GellMann:1984sj,GellMann:1985if,Chakrabarty:1989uf}. In that case, we will have a non-exponential
expansion of the ordinary space which is unacceptable.
Hence, the new term we introduced to the original model will be a hope for a successful inflation.

To have a simultaneous solution to both inflation and dynamical compactification, we expect  a cosmological
solution with the extra dimensions contract exponentially while the ordinary space expands exponentially, namely
\bea
\label{solution}
a(t) = a_0\, e^{H\,(t-t_0)} ~~;~~ b(t) = b_0\, e^{-h\,(t-t_0)}\,.
\eea
By Einstein's field equation, the various components derived from \eqref{RAB} represent the matter source that leads to
the geometry realized by the metric (\ref{metric}). Thus, we proceed to solve the equations of motion
(\ref{R00}), (\ref{Rij}) and (\ref{Rmn}) by using the geometry
of the metric (\ref{metric}). This requires matching \eqref{R00} with \eqref{R0}, \eqref{Rij} with \eqref{Ri},
\eqref{Rmn} with \eqref{Rm}, while having the solution \eqref{solution} admitted at the same time. A consistent
solution requires the metric of the target space manifold to satisfy:
\bea
\label{h0}
h_{00} &=& - \frac{M^{2+N}}{2\,N^2\, g\,\alpha^2}\,(3\,H^2+N\,h^2)\,,\\
\label{hi}
h_{ij} &=&  \frac{M^{2+N}}{2\,N^2\, g\,\alpha^2} \,(3\,H-N\,h)\,H \,a^2_0\, e^{2\,H\,(t-t_0)} \,g_{ij}\,\,, \\
\label{hm}
h_{mn} &=&  \frac{M^{2+N}}{2\,(1/\la^2+N^2\, g)\,\alpha^2} \,\left[\,-(3\,H-N\,h)\,h\, b^2_0\, e^{-2\,h\,(t-t_0)}+2\,K\,\right] g_{mn}\,.
\eea
The above expressions for $h_{AB}(z)$ verify what we have stated earlier, namely blocks of $h_{AB}(z)$ are proportional to blocks
of $G_{AB}$.

In fact, a simple inspection of the above equations reveals that we can actually further simplify our solution in a physically motivated
way. The strategy goes as follows.
Since the extra dimensions are contracting exponentially while the three ordinary dimensions are expanding
exponentially, the total spatial volume $V$ of the $3+N$ dimensional space will be given by
\bea
V \propto e^{\,(3\,H-N\,h)\,t}\,.
\eea
Imposing $3\,H-N\,h=0$ keeps the total spatial volume $V$ of the $3+N$ dimensional space constant. This means that the
exponential contraction of the extra dimensions is exactly compensated by the exponential expansion of the ordinary space.
As a consequence, inflation appears to be driven by dynamical compactification, and vice versa. Also, the condition
$3\,H-N\,h=0$ gets rid of all the time dependence in $h_{ij}$ and $h_{mn}$. The volume-preserving condition may, in some sense, be attributed to a generalized type of energy conservation. The argument goes as follows. We can parameterize the stress tensor as $T^A_{~B}= \textrm{diag}(\,\rho, \,-p, \,-p, \,-p, \,-p, ..., \,-p\, )$ where $\rho$ is the energy density and $p$ is the pressure. The covariant conservation of the stress tensor leads to $d(\rho\, a^3\, b^N\,)= - p\;d(a^3 \,b^N)$. Once we take the rate of change of energy within a comoving volume to be zero, we have $d (\rho\, a^3\, b^N)/dt =0$. This requires $d(a^3 \,b^N )/dt=0$ and hence $3\,H-N\,h=0$.

By imposing the volume-preserving
condition ($3\,H-N\,h=0$), the form of the target space manifold metric required to realize the solution \eqref{solution} is greatly
simplified, namely
\bea
\label{h00}
h_{00} &=& - \frac{M^{2+N}\,(N+3)}{6\,N\, g\,\alpha^2}\;h^2\,,\\
\label{hij}
h_{ij} &=&  0 \,,\\
\label{hmn}
h_{mn} &=&  \frac{M^{2+N}\, K}{(1/\la^2+N^2\, g)\,\alpha^2} \;g_{mn}\,.
\eea
Apart from simplicity, the time independence of $h_{AB}$ also ensures that the target space is static, despite the exponential expansion and contraction of the ordinary and extra spaces respectively.

Let us give a  physical interpretation of our results. First of all, since the non-linear sigma fields are only
dependent on the coordinates of the extra dimensions as dictated by \eqref{ansatz3}, they only exist in the extra dimensions.
However, if the volume-preserving condition is not imposed, then the non-trivial time dependence in $h_{ij}$ and $h_{mn}$ will
back-react on the time-coordinate. This means that while the ordinary 4D spacetime appears to be completely empty, the
non-linear sigma fields are affecting it indirectly. As a result, the dynamics of the non-linear sigma fields simultaneously
drive both of the exponential expansion and dynamical compactification in a non-trivial way.

On the contrary, suppose that the volume-preserving condition is imposed. As we can observe from \eqref{h00}, \eqref{hij} and
\eqref{hmn}, all of the non-trivial time dependence in $h_{ij}$ and $h_{mn}$  disappears. In this case, the non-linear sigma fields
can no longer back-react on the time-coordinate, so the ordinary 4D space-time is completely empty and free of any indirect effects. Since the non-linear sigma fields only exist in the extra dimensions, they will preferentially trigger the
dynamical compactification. The extra dimensions are rolling up to form a manifold exhibiting the same shape as that of the target space
manifold, which can be understood from \eqref{hmn}. At the same time, the exponential contraction of the extra dimensions is being exactly
compensated by the exponential expansion of the ordinary space. Therefore, the non-linear sigma fields induce the
dynamical compactification of the extra dimensions, which in turn drives inflation. No inflatons are required in the ordinary 4D spacetime.

The volume-preserving condition is physically motivated and has led to some simplifications to our solution.
When it is imposed, we can interpret inflation as driven by dynamical compactification. This means that the two physically important
processes, inflation and dynamical compactification, are both explained, and connected in a remarkable way.

Finally, we would like to provide an analysis of the equation of motion for the non-linear sigma fields in order to make sure that everything is consistent  with the anzatz invoked in \eqref{ansatz1}, \eqref{ansatz2} and \eqref{ansatz3}. The equation of motion for a fixed component field $\phi^\beta$ is given by
\bea
\label{EOM}
&& 2\,\partial_A \left(\,G^{AB}\, h_{\beta C}\,\partial_B\,\phi^C\,\right) + 2\,g\,\lambda^2\, \partial_C \left(\,G^{AB}\, h_{AB}\,\partial^C\,\phi_\beta\,\right)
\nonumber \\
&& - \frac{\partial}{\partial\,\phi^\beta}\,\left(\,G^{AB}\, h_{CD}\,\partial_A\,\phi^C\,\partial_B\,\phi^D \,\right)
- g\, \lambda^2\, \frac{\partial}{\partial\,\phi^\beta}\,\left(\,G^{AB}\, h_{AB}\,(\partial_C\,\phi^C)^2 \,\right) =0\,.
\eea
According to \eqref{ansatz1} and \eqref{ansatz2}, $\phi^0$ and $\phi^i$ are trivial fields and so they do not evolve. This is compatible with \eqref{EOM} in the sense that the equations of motion for $\phi^0$ and $\phi^i$ are not defined. For $\phi^m$, where $m=5,\,6,\,...,\,4+N$, one can easily verify that the equation of motion is satisfied by \eqref{ansatz3}, with the aid of the conditions $h_{00} \propto g_{00}$, \,$h_{ij} \propto g_{ij}$\, and\,  $h_{mn} \propto g_{mn}$ \,required by \eqref{h0}, \eqref{hi} and \eqref{hm} respectively.

\section{Discussions and Conclusions}

By considering a $4+N$ dimensional Einstein gravity coupled to a non-linear sigma model, we have been able to
provide a simultaneous solution to inflation in 4D and dynamical compactification of the extra dimensions.
The non-linear sigma fields induce the dynamical compactification, which in turn drives the inflation without inflatons.
Our solution is valid only if the number of non-linear sigma fields is exactly the same as the
dimension of the spacetime. To elaborate, we have provided a model of inflation \emph{without inflatons}, where the inflation is solely driven by the dynamical compactification of extra dimensions. The reason that we have no inflatons in our 4D spacetime is because \emph{none} of the non-linear sigma fields depends on any of our 4D spacetime coordinates. In other words, the non-linear sigma fields only live in the extra dimensions. Our 4D spacetime is completely empty until we add matter fields.

Compared to those inflationary models involving inflatons, our scheme has not improved the issue of fine-tuning as we need to start out with an action with a specific coupling between gravity and the non-linear sigma fields. As for initial singularity, the issue has been alleviated in the sense that the initial size of the spacetime does not need to be infinitely small.

One may contend that any true and physical solution to inflation
must be able to stop it at some point. How this happens in our model has yet to be determined, but we can still offer a heuristic
argument of how this may be achieved. The extra dimensions contract exponentially and will eventually reach the Planck length scale.
Since the Planck length is generally believed to be the minimum fundamental length scale in Nature, we expect
that the contraction will be forced to stop when this scale is reached. In this case, $h$ will be identically zero. Due to the
volume-preserving condition ($3\,H-N\,h=0$), the ordinary space is forced to stop expanding accordingly.

Another issue is particle production after inflation. The question is: How do we produce particles given that the 4D spacetime is
completely empty? When the extra dimensions are forced to stop contracting at the Planck length, they will undergo a period of abrupt
deceleration. This will produce a huge amount of entropy which is manifest as the production of superheavy Kaluza-Klein (KK) particles.
The natural mass scale of these KK particles is the Planck mass, and they will readily decay into all of the observed particles in
our 4D spacetime.

Of course, what we have just discussed is a possible physical scenario. For a realistic model for inflation,
we need a detailed understanding of how inflation ends and how particles are produced.
It is also imperative to show that the spectrum of density perturbations in this theory is consistent with observations.
We will explore these questions in a forthcoming article.

\begin{acknowledgments}
We thank Soo-Jong Rey, Djordje Minic and Jack Ng for useful discussions. This work  was supported by US DoE
grant DE-FG05-85ER40226.
\end{acknowledgments}

\end{document}